\documentclass[DIV=14,journal,utf8,a4paper]{scrartcl}

\usepackage[sort,compress,numbers,square]{natbib}
\usepackage{graphicx}
\usepackage{authblk}
\usepackage{listings}
\usepackage{color}
\usepackage{url}
\usepackage{relsize}
\usepackage[USenglish,english]{babel}
\usepackage[utf8]{inputenc}

\newcommand{\Rplus}{\protect\hspace{-.1em}\protect\raisebox{.25ex}{\smaller{\smaller\textbf{+}}}}
\newcommand{\Cpp}{\mbox{C\Rplus\Rplus}}
\definecolor{querybg}{gray}{0.8}

\newcommand{\query}[1]{\sloppy{\lstinline[basicstyle=\ttfamily]|#1|}}

\usepackage{scalerel}
\usepackage{tikz}
\usetikzlibrary{svg.path,calc}

\definecolor{orcidlogocol}{HTML}{A6CE39}
\tikzset{
  orcidlogo/.pic={
    \fill[orcidlogocol] svg{M256,
      128c0, 70.7-57.3, 128-128,
      128C57.3, 256, 0, 198.7, 0,
      128C0,57.3,57.3,0,128,
      0C198.7,0,256,57.3,256,128z};
    \fill[white] svg{M86.3,186.2H70.9V79.1h15.4v48.4V186.2z}
                 svg{M108.9,79.1h41.6c39.6,0,57,28.3,57,53.6c0,27.5-21.5,53.6-56.8,53.6h-41.8V79.1z M124.3,172.4h24.5c34.9,0,42.9-26.5,42.9-39.7c0-21.5-13.7-39.7-43.7-39.7h-23.7V172.4z}
                 svg{M88.7,56.8c0,5.5-4.5,10.1-10.1,10.1c-5.6,0-10.1-4.6-10.1-10.1c0-5.6,4.5-10.1,10.1-10.1C84.2,46.7,88.7,51.3,88.7,56.8z};
  }
}

\DeclareRobustCommand\orcidicon[1]{\href{https://orcid.org/#1}{\mbox{
\begin{tikzpicture}[ overlay,remember picture]
\coordinate (A);
\coordinate(B) at ($(A)+(-5pt,10pt)$);
\end{tikzpicture}
\begin{tikzpicture}[overlay,remember picture,yscale=-0.05,xscale=0.05,transform shape]
\pic at (B) {orcidlogo};
\end{tikzpicture}
}{}}}
\usepackage[colorlinks]{hyperref}

\author[1,2,*]{Timm Fitschen \orcidicon{0000-0002-4022-432X}}
\author[1,*,$\alpha$]{Alexander Schlemmer \orcidicon{0000-0003-4124-9649}}
\author[1]{Daniel Hornung \orcidicon{0000-0002-7846-6375}}
\author[1,2]{Henrik tom Wörden \orcidicon{0000-0002-5549-578X}}
\author[1,2,3]{Ulrich Parlitz \orcidicon{0000-0003-3058-1435}}
\author[1,2,3,4,5]{Stefan Luther \orcidicon{0000-0001-7214-8125}}

\affil[$\alpha$]{\small Corresponding author: alexander.schlemmer@ds.mpg.de}
\affil[*]{\small These authors contributed equally to this work.}
\affil[1]{\small Max Planck Institute for Dynamics and Self-Organization, Göttingen, Germany}
\affil[2]{\small Institute for Nonlinear Dynamics, Georg-August-Universität, Göttingen, Germany}
\affil[3]{\small German Center for Cardiovascular Research (DZHK), partner site Göttingen, Germany}
\affil[4]{\small Institute of Pharmacology and Toxicology, University Medical Center Göttingen, Göttingen, Germany}
\affil[5]{\small Department of Physics and Department of Bioengineering, Northeastern University, Boston, USA}

\title{CaosDB - Research Data Management for Complex, Changing, and Automated Research Workflows}

\begin{document}

\maketitle


\begin{abstract}
  We present CaosDB, a Research Data Management System (RDMS) designed to
  ensure seamless integration of inhomogeneous data sources and repositories of
  legacy data in a FAIR way.
  Its primary purpose is the management of data from biomedical sciences, both from simulations and experiments during the complete research data lifecycle.
An RDMS for this domain faces particular challenges:
Research data arise in huge amounts, from a wide variety of sources, and traverse a highly branched path of further processing.
To be accepted by its users, an RDMS must be built around workflows of the scientists and practices and thus support changes in workflow and data structure.
Nevertheless it should encourage and support the development and observation of standards and furthermore facilitate the automation of data acquisition and processing with specialized software.
The storage data model of an RDMS must reflect these complexities with appropriate semantics and ontologies while offering simple methods for finding, retrieving, and understanding relevant data.
We show how CaosDB responds to these challenges and give an overview of
its data model, the CaosDB Server
and its easy-to-learn CaosDB Query Language.
We briefly discuss the status of the implementation, how we currently use CaosDB, and how we plan to use and extend it.
\end{abstract}


\section{Background}
\label{sec:introduction}

Despite the technological advances over the last decades, the scientific
community still faces the problem of storing and accessing scientific data in a
structured and future-proof manner
\cite{Nelson2011,doi:10.1197/jamia.M2114,doi:10.1093/bib/bbs064,10.1007/978-3-319-11955-7_71}.
Although principles for good scientific data management have since been
formulated under the acronym FAIR~\cite{wilkinson_fair_2016} and are now widely
recognized in the community, real-life obstacles tend to prevent their
wide-spread adoption.
Especially in cross-disciplinary environments, the interaction between different user groups, e.g.\ 
numerical scientists conducting simulation studies and experimenters working in the laboratory, often leads to highly inhomogeneous approaches to data management.
For such heterogeneous data, inefficiencies become inevitable when different kinds of data have to be combined in a joint research project or when data has to be accessed by scientists who were not involved in the recording and storage procedure.
In the worst case, this can lead to data being \emph{de facto} inaccessible after their creators
can no longer be reached.

\begin{figure} 
	\centering
  \includegraphics[width=\textwidth]{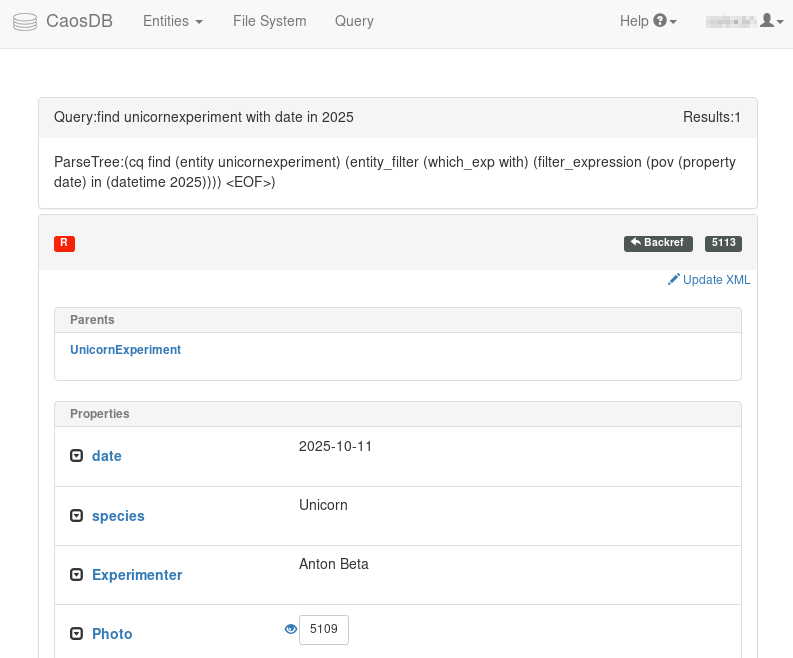}
  \caption{Screenshot of the WebUI (graphical web user interface) of CaosDB showing
  a search query that returned a fictitious experiment.}
	\label{fig:webinterface}
\end{figure}

The ongoing issues are rooted in some ubiquitous properties of scientific environments themselves:

\begin{itemize}
\item Scientists use special or customized tools, software and data formats with
  good reason. A research data management system (RDMS) must be built around
  their workflows and practices and be open for change.
\item If the system imposes too many restrictions on individual scientists they
  are likely to be unwilling or even unable to use it.
\item If the RDMS requires too much extra work for learning and understanding,
  it is likely that the individual scientist will be unable to use it
  efficiently or just be unwilling to use it at all.
  This holds in particular for the construction of queries which should retrieve
  data according to powerful criteria while being simple and intuitive at the
  same time.

\item The system should strongly encourage to develop and use standards for workflows and data models without being overly restrictive.
  Users of the database can only profit from the system when data is organized sufficiently structured to enable everyone to search and retrieve data easily and to understand the structure of the data intuitively.
  At the same time the database has to be prepared for constantly evolving data models and standards.

\item File systems and other types of storage usually organize data into some kind of hierarchy which can be folders or projects.
  In scientific environments this can raise issues, especially when data
  belongs to multiple projects or is part of cooperations.
\end{itemize}

\section{Requirements}

Based on the considerations described in the previous section, we define the
following requirements for a data management system to address the mentioned
issues.

\begin{description}
\item[Architecture] The system must be built in a client / server architecture
  for separating the high-performance workload on the database and filesystem
  from the lightweight clients.
  Create / Read / Update / Delete (CRUD) transactions on the server side must be ACID\footnote{Atomicity, Consistency, Isolation, Durability} compliant in order to keep the structure consistent at any time.
The communication API must be built around a transparent human-readable protocol with RESTful\footnote{Representational State Transfer} identifiers \cite{rest}.
This API can then be used by libraries and clients that can be integrated into
existing data management workflows.

\item[Access control, file system] Heterogenous scientific environments require
  fine-grained access-control on object level.  In order to seamlessly integrate
  into existing data acquisition and data analysis workflows the system must be
  able to incorporate an existing file system with its grown folder structure.

\item[Query language]
One of the most important requirements is the query language which has to fulfill several properties that guarantee that heterogenous data in big amounts can be searched and retrieved easily.
The logic behind the query language can also have a major impact on the data models used.
To spell this out more precisely, the data model and the query language must support:
\begin{itemize}
\item Entities with subtyping
\item User-defined n-ary relationships and properties
\item Integration of files and directories as entities
\item Native support for primitive data types which include
  several numeric data types with their physical units and uncertainties,
  standard compliant date and time values,
  booleans, strings and undefined values
\item Compound data types for lists, sets, tuples and dictionaries
\end{itemize}

\item[Extensibility]
  The system must be able to adapt to new software and hardware requirements.
  The simplest way to ensure this extensibility is to implement a server-side API for extensions and plug-ins.
\end{description}

\section{Implementation}

CaosDB~\cite{caosdb_2018} is our in-house solution for fulfilling these
conditions, to our knowledge it is currently the only existing software to
satisfy the mentioned requirements.

CaosDB is an object oriented database with a powerful query language based on
English natural language and a flexible and adaptive data model.  For example, a
typical query could look like this:

\begin{lstlisting}
SELECT flavour, rating, ingredients FROM Experiment WHICH HAS A room_temperature > 26C AND WHICH IS REFERENCED BY ExperimentSeries WHICH HAS A name LIKE *ice cream testing*
\end{lstlisting}

It also integrates efficient management of large data files directly into the core functionality
to accomodate specific requirements by the scientific users:
\begin{itemize}
\item offers a flexible data model
\item can be seamlessly integrated into existing workflows
\item allows search for values of specific fields (not just full text search),
  with automatic unit conversion and search for (back)references of linked
  objects.
\end{itemize}

\subsection{Architecture}

The software design follows a server/client architecture. The CaosDB server handles
all CRUD requests, implements consistency checks and translates the requests into
SQL commands which are redirected to the MySQL backend. It furthermore provides
a transparent layer for interactions with the file system.
The server frontend is written entirely in Java and is accessed using an RESTful
API over HTTP with XML messages.
The frontend also serves a web user interface
(WebUI, shown in Fig.~\ref{fig:webinterface}) written in
XSLT, HTML and JavaScript that can be used for browsing data and maintenance operations.

\begin{figure} 
	\centering
	\includegraphics[width=\textwidth]{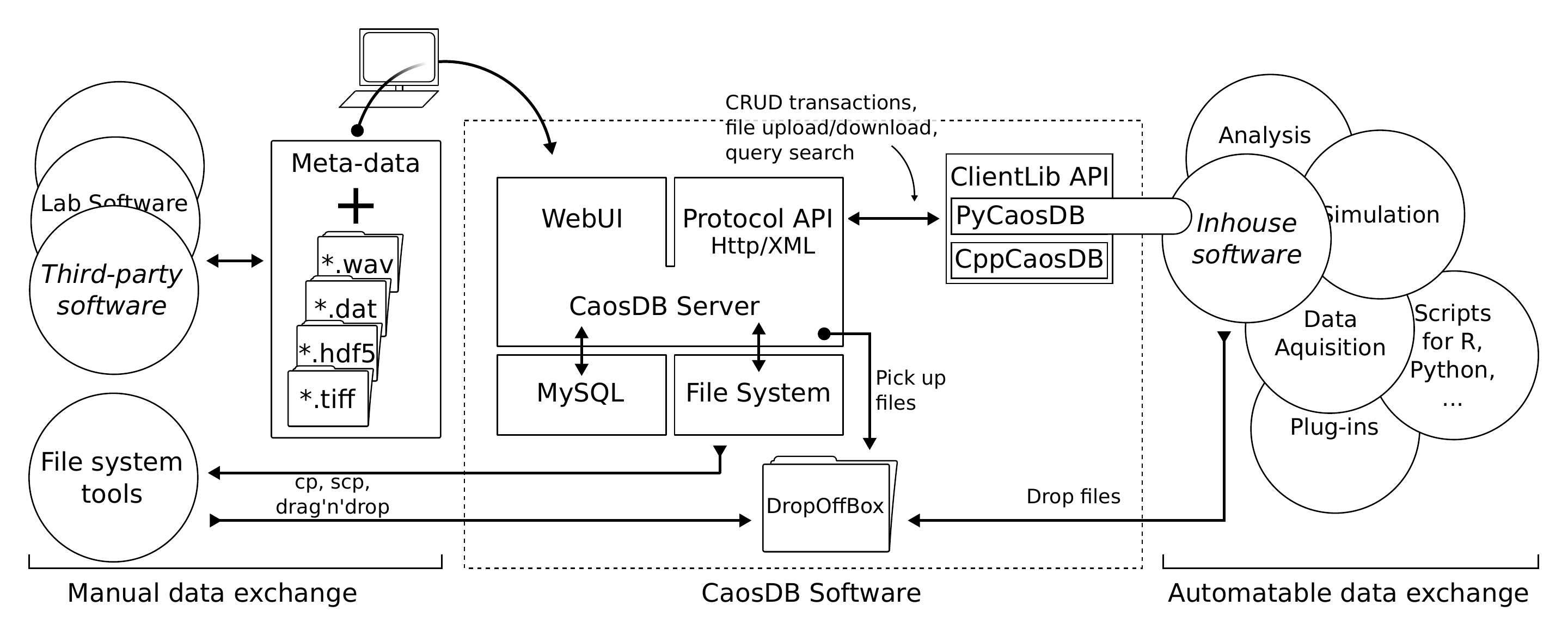}
	\caption{An overview of the components of the CaosDB project and its relations
    to contexts of application.
    The CaosDB server encapsulates access to the
    MySQL back end and the local file system.
    It implements the protocol API for
    CRUD transactions (creation, retrieval, update, deletion), file exchange,
    and query search.
    The client libraries (ClientLib API) can communicate with
    the server via the protocol API and provide interfaces in several
    programming languages for automatable data exchange with data aquisition
    software and analysis tools.
    The WebUI for convenient database access
    is directly integrated into the core application.
    These also facilitate
    the manual data exchange with non-customizable third-party tools and data
    sources.}
  \label{fig:project}
\end{figure}

The server is complemented by client libraries for Python and \Cpp{} that
encapsulate the XML API for usage in scripting, data acquisition and data
analysis tools.  Fig.~\ref{fig:project} gives a schematic overview of the
software architecture.

\subsection{Data Model}
CaosDB has a general purpose object-oriented data model which is not tied to 
any particular scientific field or structure of data.

It has a base object called \textsc{Entity}. \textsc{Entities} are
either \textsc{Record Types}, \textsc{Records}, or \textsc{Abstract Properties}
and every \textsc{Entity} has a unique, server-generated \textsc{Id}.

\textsc{Record Types} and \textsc{Abstract Properties} are used to define the ontology for a
particular domain in which the RDMS is used.
\textsc{Records} are used to store the actual data and therefore represent individuals or
particular things, e.g.\ a particular experiment, a particular time series, etc.

\textsc{Record Types} define classes or types of things, e.g.\ persons, experiments, timeseries, etc.
\textsc{Records} can be viewed as members of the class defined by its \textsc{Record Type}.
These classes can contain
\textsc{Abstract Properties} which define key-value relationships for properties of the things along with the expected data type and possibly the default unit, a default value, or a range of permitted values.
As files on the back-end file system are a major focus of this database
management system,
there is a special entity \textsc{File} that encapsulates typical file properties
like path, size and checksum.

\textsc{Entities} can be related via binary, directed, transitive \textsc{is-a}
relations which model both subtyping and instantiation, depending on the relata.
These relations construct a directed graph of the \textsc{Entities}.
If A \textsc{is-a} B we call A the child of B and B the parent of A.
No adamant restrictions are imposed on the relate of the \textsc{is-a} relation and
thus, \textsc{Entities} can be children of multiple \textsc{Entities}.

Each \textsc{Entity} has a list of \textsc{Entity Properties}, or in short just \textsc{Properties}.
An \textsc{Entity Property} is not an \textsc{Entity} of its own, but
a triple of an \textsc{Abstract Property}, a value or \textsc{Null}, and an \textsc{Importance}. 
The values can be numericals, strings, dates, any other valid value that fits into one of
several builtin data types, or, most notably, references to other
\textsc{Entities}.
The \textsc{importance} is either \textsc{obligatory}, \textsc{recommended},
\textsc{suggested}, or \textsc{fix}.
A valid child of an \textsc{Entity} implicitly inherits its parent's
\textsc{Properties} according to their \textsc{Importance}, which means that it
is obliged, recommended or only suggested to have a \textsc{Property} with the
same \textsc{Abstract Property} (or any subtype thereof).

As opposed to \textsc{Properties} with other priorities, \textsc{Fixed
  Properties} have no effect on the \textsc{Entity}'s children.
During the creation or update of \textsc{Entities}, the \textsc{importances} of
the parents are being checked by the Server.  Missing \textsc{obligatory}
\textsc{Properties} invalidate the transaction and result in an error, by
default.  Missing \textsc{Properties}, when they are \textsc{recommended},
result in a warning, but the transaction is considered valid.  Entities with
missing \textsc{suggested} \textsc{Properties} are silently accepted as valid.

This novel approach to ontology standardization
is inspired by the operators from \emph{deontic logics},
the logics of obligation and permission~\cite{Follesdal1971}.
It is designed to guide the users without restricting them too heavily and ensures that they do not insert their data wrongly \emph{by accident}.
Furthermore, it helps them to find the most relevant or best fitting \textsc{Properties} for their \textsc{Entity} based on the supertype(s).

CaosDB thus facilitates the definition and observation of standards for data storage.

\subsection{Query Language}
The CaosDB Query Language (CQL) is designed to express simple questions with simple queries resembling English. Its syntax is illustrated in Fig.~\ref{fig:ebnf} using
EBNF\footnote{Extended Backus-Naur Form}. The language is case-insensitive, but for clarity some terms are explicitely spelled in upper or mixed case here.

\begin{figure*}
    \begin{lstlisting}
cql              = query prefix, [entity type], entity name,
                   [filter separator,  filter] ;
query prefix     = "FIND" | "COUNT" | select clause ;
select clause    = "SELECT", field, {",", field}, "FROM" ;
entity type      = "ENTITY" | "RECORDTYPE" | "RECORD"
                   | "PROPERTY" | "FILE" ;
entity name      = ? any string ? ;
filter separator = "WHICH", ["HAS A"] | "WITH" ;
filter           = conjunction | disjunction | negation
                   | propery name, operator, value
                   | back-reference | ...
...
\end{lstlisting}
    \caption{
     The first levels of the CQL syntax in EBNF. This is only a
    schematic overview and does not include the syntactic sugar or white
    spaces. However, it should be noted that the top level of this syntax is
    not too complex and has only very few keywords. Yet even simple queries are
    very powerful, mainly due to the transitivity of the \textsc{is-a} relation.}
  \label{fig:ebnf}
\end{figure*}

The first term (\textsc{query prefix} in Fig.~\ref{fig:ebnf}) in a CQL
expression is the desired return type of the query:
\begin{itemize}
\item A query starting with \textsc{Count} returns a non-negative integer.
\item A query starting with \textsc{Find} returns a list of entities. 
\item A query starting with \textsc{Select} returns a table containing the values of
  selected \textsc{Properties}.
\end{itemize}

This is optionally followed by an \textsc{entity type} which restricts the query
to specific entities. The most important information searched for is probably
the \textsc{entity name} which specifies the actual ``thing'' searched for.
This term makes use of the object-oriented structure of the database and - in addition to
searching for all entities having a specific name - also returns subtypes and \textsc{Records}
being of that type.

In an CQL expression, \textsc{entity name} is followed by a list of filters
which are connected by filter separators.
Filters can address any possible \textsc{Property} of
an \textsc{Entity} and restrict the values to ranges or particular values,
use a range of comparison operators, and even search with wildcards or regular expressions.
Furthermore, relations between \textsc{Entities} can be expressed precisely.
Filters can be combined with logical operators
like \textsc{And}, \textsc{Or}, and \textsc{Not}.

The query processor is able to interpret and convert physical units.
This unique feature simplifies working with scientific data
and sets CQL apart from
SQL and various modern query languages for RDF(S), OWL or graph data.

We will illustrate the basic concepts by giving some typical examples:

\query{COUNT Experiment with date in 2017} will return the number of experiments
from 2017. In this query, \texttt{Experiment} is typically the name of a \textsc{Record
  Type} with a possibly large number of subtypes and instances. All \textsc{Entities}
which have the name \texttt{Experiment} or have a parent with this name are filtered
for those which have a \textsc{Property} with the name \texttt{date} and a date value
in the year \texttt{2017}.

CQL filters can also express the equivalence of complex SQL joins in an easily
understandable syntax:

\query{FIND Person which is referenced as an Author by an Article which has a
  Title like *terminating ventricular fibrillation*}

In this example, \texttt{Person} is a \textsc{Record Type}.  \texttt{Article} is
another \textsc{Record Type} having an \texttt{Author} and a \texttt{Title} as
\textsc{Properties}. The statement would therefore return all \textsc{Records},
if they are a \texttt{Person}, that are assigned as values of an \texttt{Author}
\textsc{Property} of a \textsc{Record} of type \texttt{Article} with a specific
title.  Since the returned objects are themselves \textsc{Records} of
\textsc{Record Type} \texttt{Person}, they have \textsc{Properties}, presumably
a name, affiliation(s), possibly an ORCiD, an email-address or some other
contact information.

Another special feature are
\textsc{Select} queries which follow an SQL-like syntax and represent their results as a table.
E.g. the result of
\query{SELECT first name, family name from person with date of birth > 2000}
will appear as an HTML table in the WebUI (downloadable as a tsv table), with
three columns
-- id, first name, and family name. This feature is intended to provide one of the interfaces
between CaosDB and existing scientific workflows.

CQL is inspired by SQL and therefore probably feels familiar to users with
knowledge of prevalent database management systems.
It should be clear from the aforementioned examples that the query language is
structured, precise and powerful, but nevertheless resembles English sentences.
This makes it easier to learn for users without SQL experience.

\subsection{User Management and Access Control}
CaosDB provides a fine-grained role-based access control system with access control lists.
It is possible to define the permissions for insertion, update, retrieval
and deletion of \textsc{Entities}, single \textsc{Properties}, and \textsc{is-a} relations, as
well as the access to the transaction log and the user management.

CaosDB has a built-in user database where users can sign up or be registered by administrators.
Furthermore, users can login with the credentials of their user
accounts from PAM (Pluggable Authentication Modules).
Access roles -- which are relevant for the authorization -- can be assigned to
clients based on various criteria including their authentication status,
the Unix groups of the user -- if PAM is used
--, and connection details, like IP address and others.
This makes it possible to share subsets of the data base with collaborators and
even a greater audience of anonymous users.

\section{Discussion}

CaosDB is currently in beta testing stage and handling around 40TiB of
experimental data
from biomedical physics in 250000 \textsc{Files} along with detailed meta data
contained in about 320 \textsc{Record Types} and 95000 \textsc{Records}.
Data file types include video recordings from optical imaging, electrophysiological time series,
scanned lab notes and image files.
Furthermore, data and parameters from simulations and information about source
code is stored along with analysis results of experimental and numerical data.
The analysis results are thereby linked to the data from which they stem.
Many file types are automatically parsed and integrated as
parts of
\textsc{Records} into our data model. The hash sums computed and stored for every
file allow for comprehensive consistency checks.
One advantage of our strict separation
between file system and data model is that the system can be directly used on top of the
established file system structure without the need to move or modify any existing file.

\begin{figure} 
  \centering
  \includegraphics[width=\textwidth]{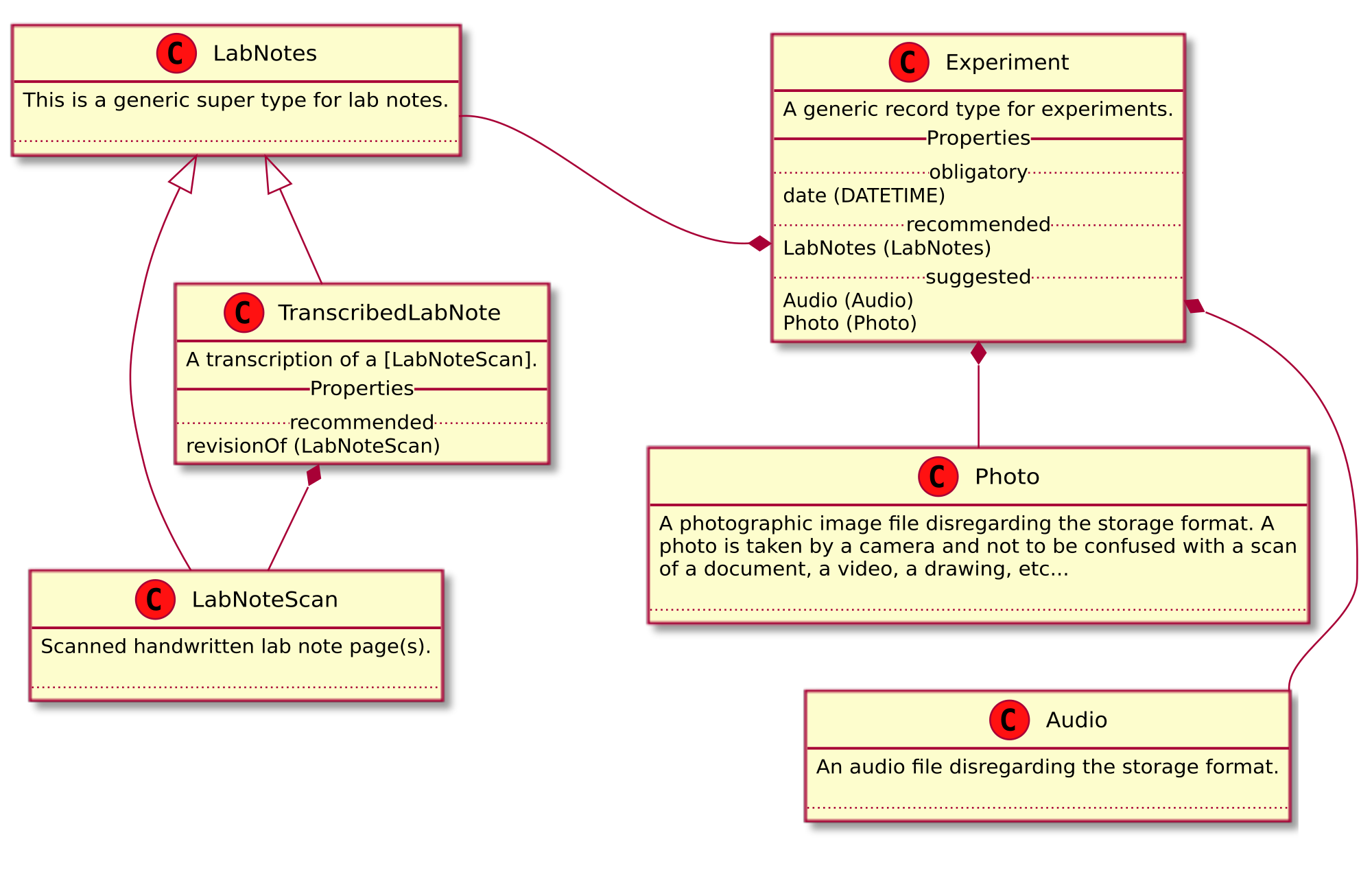}
  \caption{A UML diagram of a small excerpt of the data model we currently use
    for experimental data.}
  \label{fig:uml}
\end{figure}

For data analysis mainly the CaosDB Python client is used which can directly query
and retrieve the relevant data and use it for more specific analyses.
The power of CQL leads to much complexer SQL queries in the background than what
users would typically enter manually, and subsequently to perceived slower
responses.  Still it already proves to be faster than using simple file
operations for finding specific data.

However, the query language processing
is still subject to algorithmic optimizations.

Our current efforts to improve the data model focus on connecting experimental data
to intermediate and final results of data analysis, and the integration
of data from cardiac simulations.

The software is publicly available at \url{https://gitlab.gwdg.de/bmp-caosdb}
under the GNU Affero General Public License~\cite{agplv3}.

\section{Conclusions}

In this article we presented our approach to improve research data management in
heterogenous scientific environments.  We presented our perspective on the
current situation and proposed a list of requirements an RDMS must fulfill and
further described how our research data management system lives up to these
requirements.

The most important differences to existing solutions include a smart data model
framework which enforces the development of standards while allowing for enough
flexibility to adapt to rapidly changing scientific workflows.
Furthermore, the powerful and intuitive query language enables quick access to data
and simplifies data retrieval and data analysis.

We conclude that our database management system could provide a solution for ongoing
issues with research data management in heterogenous environments and promote
the development of standards of data storage and retrieval.  It will thereby improve
the FAIRness of research data management.

\section*{Software availability}

The public git repository with the source code is available at
\url{https://gitlab.gwdg.de/bmp-caosdb} as version \texttt{v0.1}, the program
version described here can be accessed at
\url{http://dx.doi.org/10.17617/3.1s}~\cite{caosdb_2018}.  The software
requirements are: Java 1.7 or higher, MySQL or MariaDB, Python.  The software is
licensed under the GNU~AGPLv3.

\section*{Authors' contributions}

T.F., D.H., A.S., H.t.W. are the main authors of the software, U.P. and
S.L. contributed conceptual ideas and feedback, all authors contributed to
writing the article.

The creation of a company for providing commercial support for CaosDB is planned
by: T.F., D.H., A.S., H.t.W.

\section*{Acknowledgement}

We would like to thank the contributors to the software listed in the file
\texttt{caosdb/HUMANS.md} and especially Philip Bittihn and Johannes
Schröder-Schetelig for their valuable ideas on the design of CaosDB.

We acknowledge support from the German Federal Ministry of Education and Research (BMBF) (project FKZ 031A147, GO-Bio), the German Research Foundation (DFG)  (Collaborative Research Centers SFB 1002 Project C3 and SFB 937 Project A18) and the German Center for Cardiovascular Research (DZHK e.V.).

\bibliography{references}

\end{document}